\documentclass[aps,prd,11pt,nofootinbib]{revtex4}

\usepackage[utf8]{inputenc}
\usepackage{graphicx}
\usepackage{bm}
\usepackage{amsmath,amssymb}
\usepackage{latexsym}
\usepackage{color}
\usepackage{tabularx}
\usepackage{nicematrix}
\usepackage{tikz}

\begin{document}

\title{Accurate Quasinormal Modes of the Five-Dimensional Schwarzschild-Tangherlini Black Holes}

\author{Jerzy Matyjasek}
\email{jurek@kft.umcs.lublin.pl, jirinek@gmail.com}
\affiliation{Institute of Physics,
Maria Curie-Sk\l odowska University\\
pl. Marii Curie-Sk\l odowskiej 1,
20-031 Lublin, Poland}

\begin{abstract}
The objective of  this paper is to construct the accurate (say, to 11 decimal places) 
frequencies of the quasinormal modes of the 5-dimensional Schwarzschild-Tangherlini  
black hole using three major techniques: the Hill determinant method, the continued 
fractions method and the WKB-Pad\'e method and to discuss the limitations of each. 
It is shown that for the massless scalar, gravitational tensor, gravitational vector 
and electromagnetic vector perturbations considered in this paper, the Hill determinant 
method and the method of continued fractions (both with the convergence acceleration) 
always give identical results, whereas the WKB-Pad\'e method gives the results that are amazingly 
accurate in most cases. Notable exception are the gravitational vector perturbations 
($j =2$ and $\ell = 2 $), for which the  WKB-Pad\'e approach apparently does not work.
Here we have interesting situation in which the WKB-based methods (WKB-Pad\'e and WKB-Borel-Le Roy)
give the complex frequency that differs from the from the result obtained within the framework
of the continued fraction method and the Hill determinant method. For the fundamental mode,
deviation of the real part of  frequency  from the exact value is $0.5\%$ whereas the deviation 
of the imaginary part is $2.7\%.$
For $\ell \geq 3$ the accuracy of the WKB  results is similar again to the accuracy obtained for other
perturbations. The case of the gravitational scalar perturbations is briefly discussed.

\end{abstract}
\maketitle

\section{Introduction}
\label{intro}

The quasinormal modes of the static and spherically symmetric black holes are 
the solutions of the ordinary second-order Schr\"odinger-like differential equation 
\begin{equation}
\frac{d^{2}}{dx^{2}} \psi + \left( \omega^{2} - V[r(x)]\right) \psi = 0,
\label{10} 
\end{equation}
where  $V[r(x)]$ is the potential,
 $x$ is the tortoise coordinate and 
$\psi = \psi[r(x)]$  describes the radial perturbations, satisfying 
the  purely outgoing boundary conditions at infinity and purely ingoing
at the horizon. We assume that the potential is constant as  $|x| \to \infty$ 
(the limits may be different) and has a maximum at some point  $x_{0}.$ 
The complex frequencies of the quasinormal modes, $\omega,$ the real 
part of which gives the frequency of the oscillation and the imaginary part 
describes damping of the signal, are labeled by the spin weight, $j$, 
the multipole number, $\ell,$ and the overtone number $n.$ Since their discovery, 
the quasinormal modes have been the subject of intense research and debate.
Although currently we have a good understanding of their nature and the reliable 
results at our disposal span from wormholes to black holes and from pulsars 
to analogue black holes, there are still more questions than answers.
The interested reader can consult a few excellent review 
papers~\cite{KokkotasR,BertiR,KonoplyaR,NollertR}. (See also Ref.~\cite{PaniP}).

A huge number of results that have been obtained so far can be classified with 
respect to the type and dimension of the black hole, the type of the perturbations,
adapted methods and accuracy. The more specific classification may differentiate 
between long-lived and highly-damped modes, presence of the algebraically special 
solutions and the aims of the researcher. In this paper we shall concentrate on 
the static and spherically-symmetric  black holes described by the 
Schwarzschild-Tangherlini line element
\begin{equation}
ds^2=f(r)dt^2 - f^{-1}(r)dr^2 - r^2\,d\Omega^2_{D-2},
\label{frack}
\end{equation}
with 
\begin{equation}
f(r)=1 - r^{3-D},
\label{f-function}
\end{equation}
where $D$ is the dimension. Our aim is twofold. First, we calculate the highly 
accurate (accurate to, say, 11 decimal places) complex frequencies 
of the quasinormal modes of the five-dimensional Schwarzschild black holes 
for  the massless scalar, gravitational tensor, gravitational vector and 
electromagnetic vector perturbations. In this regard it can be thought of 
as an extension of the important papers~\cite{jose_5_dim1,jose_5_dim2}. 
The methods of choice are the WKB-Pad\'e method, the Hill determinant method
and the method of continued fractions. Second, we compare the methods and  
the accuracy of the results they give. We shall accept the result only if 
at least two methods agree to the assumed accuracy. All the complex frequencies 
presented in this paper do satisfy this requirement, and in most cases there 
is full agreement of all three approaches. Additionally, each result obtained 
with the Hill determinant method, has been, in fact, obtained in the course
of the two independent calculations using the four-term and the three-term 
recurrences.  Similarly, the method of continued fraction has been implemented 
in two different ways. 

Only the last of the three methods is well-known and widely used. Indeed, 
since the publication of the Leaver's paper~\cite{Leaver1} in 1985, 
the continued fractions method (with smaller or greater modifications) 
have been successfully employed in numerous cases. Usually, it is described 
as robust, highly accurate, stable and reliable. On the other hand, the Hill 
determinant method~\cite{Pancha} is less known. It is criticized for losing 
the accuracy of the higher-order overtones. However, such behavior is
also typical for other methods. It should be emphasized that in the continued 
fractions method one can find asymptotic expression describing the tail
(i.e., the remaining part of the infinite continued fraction) that certainly 
improves the quality of approximants. All this reduced the Hill determinant
method to the inferior role of supplier of the initial values of $\omega$ for
more mature methods. In this paper however, we shall propose 
an extension of the Hill determinant method, which, in our opinion, 
partially answers the criticism pronounced by some authors.

Finally, the WKB-Pad\'e method~\cite{jaOp,jago,ja2020b} is a recent extension and modification 
of the Iyer-Will approach~\cite{wkb1} (and its further generalizations~\cite{Konoplya6}).
It differs from the previously discussed methods in several aspects, the most 
important of which is its  `black box' nature with the potential $V(x)$ treated as an input 
and the accurate quasinormal modes as the output. As in other WKB-based methods, 
the frequency of the quasinormal modee is (formally) given by a series of terms constructed, 
for a given $j$ and $l,$ solely from the derivatives of the potential $V(x)$
at $x_{0}.$ Since this series is divergent, one can try to use some standard techniques 
to obtain sensible results. For this purpose, it has been proposed to apply the Pad\'e transform to the
the truncated series. This suggests that the WKB-Pad\'e method could be, in principle, applicable 
to a wide class of potentials. Indeed, it has been shown that in many cases
this procedure, despite its simplicity, yields amazingly accurate results. 
For example, and it is really impressive, all three methods give 
(to 32 digits accuracy) exactly the same frequency for the lowest fundamental 
mode of the gravitational perturbations of the Schwarzschld black hole.
\begin{equation}
 \omega = 0.74734336883608367158698400595410 -
 0.17792463137787139656092185436905 i.
 \label{acc}
\end{equation}
Moreover, to 28 digits accuracy this result is confirmed by independent 
calculations employing the confluent Heun functions~\cite{Fiziev2}. 

The literature on the quasinormal modes of the higher-dimensional 
Schwarzschild-Tangherlini black hole is quite rich. For problems 
not covered here the interested reader may consult,  for example,
Refs.~\cite{Birmingham,KonoplyaHD,Rostworowski,Berti,Soda} and the references cited therein.
Specifically, we shall not discuss behavior of the quasinormal modes
which can easily be inferred form low accuracy calculations. The ideal 
situation we have in mind is the following: Suppose that we have  very accurate measurements
of the quasinormal frequencies $\omega$ and our task is to choose between the black hole models.
From the astrophysical point of view the most important ones are the long lived modes, 
such as considered here.
Our second goal, beyond natural curiosity, is to compare accuracy and overall performance
of various competing methods.
Finally, it should be emphasized that in order to understand its pros and cons,  
every new method should be extensively tested against the existing accurate results.
Consequently the role played by the highly accurate values of the quasinormal frequencies
should not be underestimated.

The paper is organized as follows. In Sec.~\ref{fst} we provide basic equations 
and introduce the calculational strategies. Specifically, in Sec.~\ref{HDm} 
we discuss the Hill determinant method. To the author's knowledge it is a first 
attempt to improve quality of the method by using the series acceleration techniques 
in the construction of the complex frequencies of the quasinormal modes 
and a first attempt to employ the Hill determinant method for the perturbations 
of the higher-dimensional black holes. We also show, that the Gauss elimination
is not necessary ingredient for the calculations involving 4-term recurrence
relations. (See also Ref.~\cite{Leaver2}). In Sec.~\ref{CFm} we introduce 
the continued fraction method with the series acceleration and in Sec.~\ref{wkpb}
we briefly discuss the WKB-Pad\'e method. In Sec.~\ref{results1} we present the results 
of the calculation of the quasinormal frequencies of  massless scalar and gravitational 
tensor perturbations. We follow the normalization used in Ref.~\cite{jose_5_dim1}.
Similarly, in Sec.~\ref{results2} we discuss our results for gravitational 
and electromagnetic vector perturbations. Finally, in Sec.~\ref{final} 
we briefly discuss our preliminary calculations of the quasinormal frequencies 
of the gravitational scalar perturbations.

\section{The quasinormal frequencies of 5-dimensional Schwarzschild-Tangherlini black holes}
\label{fst}

The differential (master) equation describing the massless scalar, gravitational 
tensor, gravitational vector and electromagnetic vector perturbations of the 
$D$-dimensional Schwarzschild-Tangherlini black hole~\cite{Cardoso,Kodama1,Kodama2,Hartnoll}
($D>4$) can be written in a compact form:
\begin{equation}
\frac {d^2} {dx^2} \psi + \left\{\omega^{2}  - f(r) \left[ \frac {l(l+D-3)} {r^2} +
\frac {(D-2)(D-4)} {4r^2} + \frac {(1-j^2)(D-2)^2} {4r^{D-1}} \right]\right\}\psi,
\label{master}
\end{equation}
where $j$ is given by
\begin{equation}
j=
\begin{cases}
0, &       \text{massless scalar and gravitational tensor perturbations}.  \\ 
2, &       \text{gravitational vector perturbations},\\
\frac{2}{D-2}, &   \text{electromagnetic vector  perturbations}.\\
\end{cases}
\label{ma_eq}
\end{equation}
On the other hand, the potential of the gravitational scalar perturbations is more 
complicated
\begin{eqnarray}
 V = \frac{f(r)Q(r)}{4 r^2 \left[ 2 m+d\,(d+1)\xi\right]^2} \,,
\label{grav_scalar}
 \end{eqnarray}
where
\begin{eqnarray}
   Q(r) &=&  d^4(d+1)^2\xi^3
            + d\,(d+1)\left[4(2d^2-3d+4)\,m+d\,(d-2)(d-4)(d+1) \right]\xi^2
\nonumber \\
        &{}& \,
            -12d\left[(d-4)\,m+d\,(d+1)(d-2) \right] m\, \xi +16\,m^3+4d\,(d+2)\,m^2 ,
\end{eqnarray}
 $d = D-2,$ $m = l(l+d-1)-d, $  $\xi = r^{3-D}.$ 
It does not belong to the class of potentials described by Eq.~(\ref{master}),
and as such it will not be considered here in any depth. Only a few preliminary 
results obtained within the framework of the WKB-Pad\'e method will be briefly
discussed at the end of this paper. This case certainly deserves  a separate 
and more thorough study.

Following~\cite{jose_5_dim1}, we shall therefore confine ourselves to the perturbations 
characterized by $j = 0, 2/3$ and $2.$ Since one of the aims of this paper is to perform 
the stress tests of the three methods briefly discussed in Introduction one can also 
use some `unphysical' values of $j.$
For the even- dimensional black hole the perturbation function $\psi(r)$ can be expanded
as a power series
\begin{equation}
\psi(r) = \left(\frac {r-1} {r}\right)^{-i\omega /(D-3)}e^{i\omega r} \sum _{n=0}^{\infty} a_n \left(\frac {r-1} {r} \right)^n,
\label{psi_even}
\end{equation}
whereas for odd $D$ it can be expanded as
\begin{equation}
\psi(r) = \left(\frac {r-1} {r+1}\right)^{-i \omega/(D-3)}e^{i\omega r} \sum _{n=0}^{\infty} a_n \left(\frac {r-1} {r} \right)^n .
\label{psi_odd}
\end{equation}
 Upon substituting the expansions (\ref{psi_even}) and (\ref{psi_odd}) into the master equation (\ref{master})
one obtains a $(2D-5)$-term recurrence relation and  $(2D-6)$-term recurrence relation, respectively.

Now, let us confine to the five-dimensional case.
Making use of (\ref{psi_odd}), after some algebra, one obtains a four-term recurrence relations
\begin{eqnarray}
 &&  0 = \alpha_{0} a_{1} + \beta_{0} a_{0}, \nonumber \\
 &&  0 = \alpha_{1} a_{2} + \beta_{1} a_{1} + \gamma_{1} a_{0}, \nonumber \\
 &&  0 = \alpha_{k} a_{k+1} + \beta_{k} a_{k} + \gamma_{k} a_{k-1} + \delta_{k} a_{k-2},
 \label{rec_1}
\end{eqnarray}
where
\begin{eqnarray}
 \alpha_{k} &=& -2 (k+1) (k-2 \rho +1),\nonumber \\
 \beta_{k}  &=& 5 k^2+k (5-16 \rho )+  l(l+2) +16 \rho ^2-8 \rho +\frac{9 }{4}  (1-j^{2})  +\frac{3}{4},\nonumber \\
 \gamma_{k} &=& -4 k^2+8 k \rho -\frac{9 }{2} (1-j^{2} ) +4,  \\
 \delta_{k} &=& k^2-k+\frac{9 }{4} (1-j^{2}) -2
 \label{rec-coeff}
\end{eqnarray}
and $\rho = i \omega/2.$
It should be noted that our recurrence relations differ from these presented in Ref.~\cite{jose_5_dim1}
and can easily be constructed or checked with the aid of any computer algebra system. 

\subsection{The Hill determinant method}
\label{HDm}

Inspection of the recurrence relations (\ref{rec_1}) shows that
they define a sparse banded matrix $\mathcal{H}$ of the width $w=4$
\begin{equation} \mathcal{H} = 
\begin{bmatrix}
 \beta_{0}  & \alpha_{0}  &              &              &              &            &              &                &              &               &              &   \\ 
 \gamma_{1} & \beta_{1}   & \alpha_{1}   &              &              &            &              &                &              &               &              &   \\
 \delta_{2} & \gamma_{2}  & \beta_{2}    & \alpha_{2}   &              &            &              &                &              &               &              &   \\
            & \delta_{3}  &  \gamma_{3}  & \beta_{3}    &   \alpha_{3} &            &              &                &              &               &              &   \\
            &             &              &              &              & \ddots     &              &                &              &               &              &   \\ 
            &             &              &              &              &            & \delta_{n-1} & \gamma_{n-1}   & \beta_{n-1}  & \alpha_{n-1}  &              &   \\
            &             &              &              &              &            &              & \delta_{n}     & \gamma_{n}   & \beta_{n}     & \alpha_{n}   &   \\
            &             &              &              &              &            &              &                &              &               &              &   \ddots 
\end{bmatrix}
\end{equation}
The condition that a nontrivial solution of the recurrence exists is given by the equation
\begin{equation}
 \det \mathcal{H} = 0.
\end{equation}
Now, let $h_{k}$ denotes determinant of $(k+1)\times(k+1)$ 
matrix constructed from the infinite matrix $\mathcal{H}.$
Calculations of the determinants of such matrices using standard techniques 
may be impractical. The special form of $\mathcal{H}$ suggests
another approach. Indeed, the Laplace expansion along $n$-th 
row gives simple recurrence formula
\begin{equation}
 h_{n} = \beta_{n} h_{n-1} -\gamma_{n} \alpha_{n-1} h_{n-2} + \delta_{n} \alpha_{n-1} \alpha_{n-2} h_{n-3}.
 \label{rec_hill_1}
\end{equation}
Although Eq.~(\ref{rec_hill_1}) alone is sufficient for determination of the quasinormal modes, 
having in mind other applications,
we transform the four-term recurrence relations to the three-term ones.
It can easily be done with the aid of the Gauss elimination method. Standard
manipulations give:
\begin{gather}
 \alpha'_{0} a_{1} + \beta'_{0} a_{0}  =  0, \nonumber \\
  \alpha'_{k} a_{k+1} + \beta'_{k} a_{k} + \gamma'_{k} a_{k-1} =  0,
\end{gather}
where the primed coefficients are given by
\begin{equation}
 \alpha'_{k} = \alpha_{k},
\end{equation}
\begin{eqnarray}
 \beta'_{k} &=& \beta_{k}, \nonumber \\
 \beta'_{k} & = & \beta_{k} - \frac{\delta_{k}}{\gamma'_{k-1}} \alpha'_{k-1} ,
\end{eqnarray}
\begin{eqnarray}
  \gamma'_{k} &=& \gamma_{k} ,\nonumber \\
 \gamma '_{k} & = & \gamma_{k} - \frac{\delta_{k}}{\gamma'_{k-1}} \beta'_{k-1} .
\end{eqnarray}
The thus obtained three-term recurrence  defines a tridiagonal matrix $\mathcal{H}'$ (a sparse banded matrix
of width 3)
\begin{equation} \mathcal{H'} = 
\begin{bmatrix}
 \beta'_{0} & \alpha'_{0} &              &            &             &                &             &                &              &\\    
 \gamma'_{1}& \beta'_{1}  & \alpha'_{1}  &            &             &                &             &                &              &\\
            & \gamma'_{2} & \beta'_{2}   & \alpha'_{2}&             &                &             &                &              &\\
            &             &              &            &   \ddots    &                &             &                &              &\\  
            &             &              &            &             & \gamma'_{n-1}  & \beta'_{n-1}& \alpha'_{n-1}  &              &\\
            &             &              &            &             &                & \gamma'_{n} & \beta'_{n}     & \alpha'_{n}  &\\
            &             &              &            &             &                &             &                &              & \ddots  
\end{bmatrix}
\end{equation}
Denoting the determinant of $(k+1)\times (k+1)$ matrix by $h'_{k},$
one has the following simple relation
\begin{equation}
 h'_{n} = \beta'_{n} h'_{n-1} -\gamma'_{n} \alpha'_{n-1} h'_{n-2},
\end{equation}
which (formally) follows from  the previous one by putting $\delta_{k} =0.$

The general idea of the calculations is as quite simple:
We truncate the series expansion at some $n,$
and calculate the determinant, which is a polynomial $(w=4)$
or a rational function $(w= 3)$ of $\rho.$ Subsequently, we find the roots, 
and, finally, identify complex frequencies 
of interest. More precisely, our strategy (the same in both cases) 
is as follows: First, we calculate the roots of the polynomials $p_{n}$
for $1\leq n\leq N$ and  identify the stable ones. With 
increasing $n$ the roots migrate on a complex plane and 
we consider the root as stable, if its location does not 
change with increasing $n$ to the precision assumed. 
Sometimes, as we shall see,  the roots approach their 
limit in a quite interesting way. Finally, we accelerate 
convergence of the series of the root approximants using 
the well-known Wynn's $\epsilon$ algorithm~\cite{wynn1}. In the case 
in hand, we assume $N = 250$  for both the  four-term and 
the three-term recurrences. Since the calculations of the quasinormal modes 
require knowlegde of \emph{all} roots of some high-order polynomial one has 
to use effective and reliable algorithms. Here we have used Jenkins-Traub
algorithm and some frequencies has been checked with the aid of the modified
Sch\"onhage algorithm \footnote{The Jenkins-Traub algorithm is default method
in Mathematica whereas the modified Sch\"onhage algorithm has been implemented in PARI/GP.}.

\subsection{The continued fractions method}
\label{CFm}

As is well known, every three-term recurrence is closely related to some
continued fraction. The standard reference for computationally oriented
research is the article by Gautschi~\cite{gautschi}. Let us recall a few 
basic facts. The three-term recurrence has generally two independent
solutions, and their particular linear combinations, $q_{k},$  with 
the property that $q_{k}/y_{k}\to 0 $ as $k \to \infty,$ where  $y_{k}$ 
is any solution not proportional to $q_{k},$ form a one-dimensional subspace
in the space of all solutions. Elements of this one-dimensional subspace 
are called minimal (minimal at infinity) and the minimal solution is 
completely determined by (one) initial value. This is very important 
because  the quasinormal mode corresponds to the minimal solution 
of a recurrence relations. The convergence condition for the series 
expansion (which simultaneously is the condition for the quasinormal modes) 
can be written in the form of the infinite continued fraction
\begin{equation}
 \beta'_{0} - \dfrac{\alpha'_{0} \gamma'_{1}}
                    {\beta'_{1} - \dfrac{\alpha'_{1} \gamma'_{2}}
                                       {\beta'_{2} - \dfrac{\alpha'_{2} \gamma'_{3}}{\beta'_{3} - \dots}}} = 0.
                                       \label{cont1}
\end{equation}
In what follows we shall use the more popular notation, in which 
the above equation can be written in the form
\begin{equation}
 \beta_{0} - \frac{\alpha_{0} \,\gamma_{1}}{\beta_{1} -}\, \frac{\alpha_{1}\, 
 \gamma_{2}}{\beta_{2} -}\, \frac{\alpha_{2}\, \gamma_{3}}{\beta_{3} -}\dots = 0.
 \label{cont2}
\end{equation}
Inverting Eq.~(\ref{cont1}) $n$ times, one obtains
\begin{equation}
 \beta_{n} -\frac{\alpha_{n-1} \,\gamma_{n}}{\beta_{n-1} -}\,\frac{\alpha_{n-2} \,
 \gamma_{n-1}}{\beta_{n-2} -}\,\dots -\frac{\alpha_{0} \gamma_{1}}{\beta_{0}} =
 \frac{\alpha_{n} \,\gamma_{n+1}}{\beta_{n+1} -}\, \frac{\alpha_{n+1}\, 
 \gamma_{n+2}}{\beta_{n+2} -}\, \frac{\alpha_{n+2}\, \gamma_{n+3}}{\beta_{n+3} -}\dots.
 \label{cont3}
\end{equation}
Both forms are equivalent and may serve as defining equations for calculations 
of the quasinormal frequancies. Now, our strategy is as follows. First, we 
generate successive the approximants of the continued fractions up to some
$N$ (in our case $n \leq 250$) and solve the thus obtained  equations. Each 
approximant is some function of $\omega.$ In the next step we identify its stable 
roots and accelerate convergence of the series of successive approximations to $\omega$ using the $\epsilon$ 
algorithm. In order to make some independent checks we have used both (\ref{cont2}) 
and (\ref{cont3}) and since we are interested in a fundamental modes
and their few long-lived overtones, only  basic estimations of a tail 
have been implemented. The more sophisticated calculations would require construction
of the asymptotic representation of the remainder of the infinite continued fraction.
For the three-term recurrence construction of such asymptotic formula is 
a five-finger exercise~\cite{Leaver1,Nollert2}, however, as the number 
of consecutive Gauss eliminations necessary to construct the three-term recurrence
grows with the dimension $D$ this problem becomes harder and harder~\cite{Leaver2}.
This may lead to the need to modify the computational strategy.

\subsection{ The WKB-Pad\'e approximation}
\label{wkpb}

The  WKB-based methods are very popular tools for calculating the frequencies 
of the quasinormal modes. They are simple, yield reasonable results and depend 
 solely on the derivatives of the potential at its (global) maximum. On the 
other hand however,  they have severe limitations. Indeed, they cannot be used
in calculations of the frequencies of  higher overtones and may lead 
to erroneous results for the potentials 
which are not positive-definite or described by a complicated functions 
of the radial coordinate with additional maxima and minima~\cite{price}. 
The main idea is simple, and, effectively, it is encapsulated in the formula 
\begin{equation}
 \frac{ i Q_{0}}   {\sqrt{  2Q''_{0}}} -\sum_{k=2}^{N} 
 \Lambda_{k}= n+ \frac{1}{2},
 \label{master2}
\end{equation} 
relating $\omega,$ the overtone number $n$ and the functions $\Lambda_{k}.$ 
Each $\Lambda_{k}$ is a combination 
of the derivatives of $Q(x)= \omega^{2} -V(x)$ calculated at $x=x_{0}$ and 
its complexity grows fast with the order.  Setting all $\Lambda_{k}$ equal 
to 0 results in the Schutz-Will formula~\cite{wkb0,Bahram}, extensively used in 
the current literature to study large $\ell$ behavior of the quasinormal 
frequencies. The Schutz-Will formula is the starting point for various generalizations
and plays important role
in determining the order of magnitude and the general behavior of the modes.
Retaining $\Lambda_{2}$ and $\Lambda_{3}$ in~(\ref{master2}) 
gives famous Iyer-Will approximation~\cite{wkb1}. Typically, it yields substantially
better results that the previous one. The Iyer-Will result has been extended 
by Konoplya~\cite{Konoplya6} to include the terms up to $N = 6.$ This method usually 
gives even more accurate results then the Iyer-Will method, and it seems that 
it is, in a sense, the optimal one. Due to its simplicity and the quality 
of the results it gives it is the method of choice in a numerous applications.

The general form of the functions $\Lambda_{k}$ are known for $k\leq 16$ 
and one can easily incorporate them into the general formula~(\ref{master2}). 
It should be noted however, that since the discussed methods rely on summing 
up the $\Lambda$ terms they cannot be used to obtain highly accurate complex 
frequencies. Moreover,  increasing the number of $\Lambda$ terms does not 
improve the quality of the approximation. On the contrary, it can be shown 
that $\Re(\omega)$  and $|\Im(\omega)|$  rapidly grow with the number
of the terms of  WKB series summed. 

Our approach, consists of treating the right hand side of the expression
\begin{equation}
 \omega^{2} = V(x_{0})-i \left(n+\frac{1}{2} \right)  \sqrt{  2Q''_{0}} \tilde{\varepsilon} -i 
 \sqrt{  2Q''_{0}} \sum_{i=2}^{N} \tilde{\varepsilon}^{j} \Lambda_{j}
 \equiv V(x_{0})  + \sum_{i=1}^{N} \tilde{\varepsilon}^{i} \tilde{\Lambda}_{i}
 \label{omm}
\end{equation}
as the power series and instead of summing the terms (which is probably a bad strategy 
for higher-order calculations given its divergent nature) we construct the Pad\'e approximants~\cite{jaOp,jago}. 
As is well known, the Pad\'e approximants of a truncated power series 
$\sum a_{k} \tilde{\varepsilon}^{k}$ are defined 
as the unique rational functions $\mathcal{P}_{N}^{M}(\tilde{\varepsilon})$ 
of the degree $N$ in the denominator and $M$ in the numerator, 
\begin{equation}
 \mathcal{P}_{N}^{M}(\tilde{\varepsilon}) = \frac{\sum_{k=0}^{M} A_{k} \tilde{\varepsilon}^{k}}{\sum_{k=0}^{N} B_{k} \tilde{\varepsilon}^{k}},
\end{equation}
satisfying simple relation~\cite{CarlB}
\begin{equation}
 \mathcal{P}_{N}^{M}(\tilde{\varepsilon}) - \sum_{k=0}^{M+N} a_{k}\tilde{\varepsilon}^{k}
 ={ \cal{O}}(\tilde{\varepsilon}^{M+N+1}).
 \label{pade}
\end{equation}
Without loss of generality one can put  $B_{0}=1.$
It has been explicitly demonstrated (see Refs.~\cite{jaOp,jago,ja2020b}) that in a number of cases this  strategy yields amazingly 
accurate results. The Pad\'e summation of the WKB terms in Eq.(\ref{omm}) 
has been proposed in Ref.~\cite{jaOp} and subsequently extended in Ref.\cite{jago} 
to which the interested reader is referred  for the technical details and a general 
discussion. Although the functions $\Lambda_{k}$ for $k \geq 17$ are unknown, they 
can easily be constructed for a given potential with prescribed $\ell$ and $n$ 
numerically~\cite{jago,hatsuda,sulejman}. Finally, observe that instead of the Pad\'e 
approximants, one can apply the Wynn's algorithm to the partial sums 
of the series~(\ref{omm}).

\section{Results}
\subsection{Massless scalar and gravitational tensor perturbations}
\label{results1}

We have used all three methods to calculate the complex frequencies of the 
quasinormal modes of the massless scalar and gravitational tensor perturbations
$(j = 0)$ for $\ell = 0, 1, 2$ and their first four overtones (Tables~\ref{tab1}-\ref{tab3}). 
Although the calculations have been carried out with a very high precision,  
the results presented here are rounded to 11 decimal places. Taking $N = 250$ both the continued-fraction 
method and the Hill determinant method with the convergence acceleration give
identical result. (Note that the meaning of $N$ in each method is different).
In fact,  each result displayed in the second column of Tables~\ref{tab1}-\ref{tab3}
has been obtained using four methods: two independent calculations 
making use of the determinant of the four-diagonal and the  tridiagonal matrices
and two independent calculations of the continued fractions given by Eqs. (\ref{cont2}) 
and (\ref{cont3}), respectively. Our general calculational strategy is as follows: as long as possible we try to 
perform analytical calculations and apply the exact arithmetic. This pays off in the 
high-precision stage of the numerical calculations. 

Our results (when rounded to 4 decimal places) are either in perfect agreement or are very 
close to the results presented in 
Ref.~\cite{jose_5_dim1}. As expected, the WKB-Pad\'e method yields results that are slightly 
less accurate for the overtones of the fundamental mode, however, it gets progressively 
better with increase of $\ell.$ For example for $\ell \geq 2$ we have a perfect agreement 
of the methods. The order of the (diagonal) Pad\'e transforms, $\mathcal{P}_{k}^{k},$ 
depends on the mode and to carry out the calculations in a reasonable time $k$ never 
exceeds 300. To simplify calculations we have not attempted to look for smallest optimal 
order of the Pad\'e transforms that guarantees prescribed accuracy, instead, 
we have assumed some safe, sufficiently big, value.

Now, let us define deviations of the WKB-Pad\'e results from the results obtained within
the framework of the Hill determinant method or the method of continued fractions:
\begin{equation}
 \Delta_{R}\omega = \frac{\Re(\omega_{WKB}) -\Re(\omega_{HD})}{\Re(\omega_{HD})} 100\%
\end{equation}
and 
\begin{equation}
 \Delta_{I}\omega = \frac{\Im(\omega_{WKB}) -\Im(\omega_{HD})}{\Im(\omega_{HD})} 100\%.
\end{equation}
A closer examination of Tab.~\ref{tab1} shows that the deviations of the real part of the frequency
never exceed $7\times 10^{-3} \%$ and $7\times 10^{-7}\% $ for the real and the imaginary part,
respectively.  It is an amazing result, which could possibly be made even better by increasing the order of the 
WKB series and calculating the higher order Pad\'e transforms.
Inspection of the Tabs~\ref{tab2} and~\ref{tab3} shows, as expected, that the WKB-Pad\'e method 
gets progressively better with increase of $\ell.$ For the fundamental mode $\ell = 1$ and its 
four first overtones the accuracy of the WKB-Pad\'e based calculations is better than $1.82 \times 10^{-6}\%$ 
and $1.16 \times 10^{-7} \%$ for the real and the imaginary part of the frequencies, respectively.
For $\ell =2$ we have a perfect agreement of all three methods.

\begin{center}
\begin{table}
\caption{\label{tab1} The frequencies of the quasinormal modes of the scalar 
and gravitational tensor perturbations $(\ell = 0)$
of the five-dimensional Schwarzschild-Tangherlini black hole calculated for 
$n = 0,1,2,3,4.$ The Hill determinant method (HD) and the continued-fractions
method (CF), both with the convergence acceleration,  yield identical results. 
The WKB-Pad\'e results are slightly less accurate. The frequencies are defined as 
$\tilde{\omega} = \omega/T_{H},$
where the Hawking temperature $T_{H} = 1/ 2 \pi,$ and the last column gives 
the maximal order of the (diagonal) Pad\'e approximants.}
\begin{tabular}{|c| c | c| c| }
\hline\hline
$n $ & $\tilde{\omega}_{CF}/\tilde{\omega}_{HD}$  & $\tilde{\omega}_{WKB}$ & Pad\'e  \\ \hline
0  &  $ \underline{3.35418783669} -  \underline{2.40881848257} i $ & $ \underline{3.35418783669} -  \underline{2.40881848257} i$ &  (100,100)     \\
1  &  $ \underline{2.336462592}25 - \underline{ 8.31019918}700 i $ & $ \underline{2.336462592}53 - \underline{ 8.31019918}658 i$ &  (250,250)     \\
2  &  $ \underline{1.88699}812446 - \underline{14.786241}49094 i $ & $ \underline{1.88699}618127 - \underline{14.786241}57678 i$ &  (250,250)     \\
3  &  $ \underline{1.692}59527010 - \underline{21.2198}1386107 i $ & $ \underline{1.692}60763003 - \underline{21.2198}0227040 i$ &  (350,350)     \\
4  &  $ \underline{1.584}14637600 - \underline{27.606843}13053 i $ & $ \underline{1.584}03980753 - \underline{27.606843}06759 i$ &  (400,400)    \\
\hline\hline
\end{tabular}
\end{table}
\end{center}

\begin{center}
\begin{table}
 \caption{\label{tab2}   The frequencies of the quasinormal modes of the scalar 
and gravitational tensor perturbations $(\ell = 1)$
of the five-dimensional Schwarzschild-Tangherlini black hole calculated for 
$n = 0,1,2,3,4.$ The Hill determinant method (HD) and the continued-fractions
method (CF), both with the convergence acceleration,  yield identical results. 
The WKB-Pad\'e results are slightly less accurate. The frequencies are defined as 
$\tilde{\omega} = \omega/T_{H},$
where the Hawking temperature $T_{H} = 1/2 \pi,$ and the last column gives 
the maximal order of the (diagonal) Pad\'e approximants.}
\begin{tabular}{|c| c | c| c| }
\hline\hline
$n$ & $\tilde{\omega}_{CF}/\tilde{\omega}_{HD}$  & $\tilde{\omega}_{WKB}$ &Pad\'e  \\ \hline
0  &  $ \underline{6.38382253011}  -  \underline{ 2.27657411582} i $ & $ \underline{6.38382253011} -    \underline{2.27657411582} i$ &  (100,100)     \\
1  &  $ \underline{5.38079295983}  -   \underline{7.27345089157} i $ & $ \underline{5.38079295983} -    \underline{7.27345089157} i$ &  (150,150)     \\
2  &  $ \underline{4.16851774973}  -  \underline{13.25239223151} i $ & $ \underline{4.16851774973} -   \underline{13.25239223151} i$ &  (200,200)     \\
3  &  $ \underline{3.40134506}823  - \underline{19.708451266}83 i $ & $ \underline{3.40134506}711 -   \underline{19.708451266}19 i$ &  (300,300)     \\
4  &  $ \underline{2.9541528}7238  -  \underline{26.2148370}3856 i $ & $ \underline{2.9541528}1867 -   \underline{26.2148370}6910  i$ &  (350,350)     \\
\hline\hline
\end{tabular}
\end{table}
\end{center}

\begin{center}
\begin{table}
 \caption{\label{tab3}  The frequencies of the quasinormal modes of the scalar 
and gravitational tensor perturbations $(\ell = 2)$
of the five-dimensional Schwarzschild-Tangherlini black hole calculated for 
$n = 0,1,2,3,4.$ The Hill determinant method (HD), the continued-fractions
method (CF), both with the convergence acceleration,  and the WKB-Pad\'e 
method yield identical results. 
The frequencies are defined as $\tilde{\omega} = \omega/T_{H},$
where the Hawking temperature $T_{H} =1/ 2 \pi,$ and the last column gives 
the maximal order of the (diagonal) Pad\'e approximants.}
\begin{tabular}{|c| c | c| c| }
\hline\hline
$n$ & $\tilde{\omega}_{CF}/\tilde{\omega}_{HD}$  & $\tilde{\omega}_{WKB}$ & Pad\'e  \\ \hline
0  &  $ \underline{ 9.49117521848} -  \underline{2.24647282923} i $ & $ \underline{9.49117521848} - \underline{ 2.24647282923 }i$ &  (100,100)     \\
1  &  $ \underline{8.75072206831} -  \underline{6.94013514280} i $ & $ \underline{8.75072206831} -  \underline{6.94013514280} i$ &  (150,150)     \\
2  &  $ \underline{7.50107062040} - \underline{12.22543669696} i $ & $ \underline{7.50107062040} - \underline{12.22543669696} i$ &  (200,200)     \\
3  &  $ \underline{6.24820058005} - \underline{18.21482743438} i $ & $ \underline{6.24820058005} - \underline{18.21482743438} i$ &  (250,250)     \\
4  &  $ \underline{5.31531126304} - \underline{24.59655935942} i $ & $ \underline{5.31531126304} - \underline{24.59655935942} i$ &  (300,300)     \\
\hline\hline
\end{tabular}
\end{table}
\end{center}

\subsection{The gravitational vector and electromagnetic vector perturbations}
\label{results2}

Before we start analysis of the gravitational vector perturbations 
let us discuss a typical behavior of the Pad\'e approximants of the 
series given by Eq.~(\ref{omm}). A more thorough analysis can be found 
in Ref.~\cite{jago}. First, observe that for sufficiently large $\ell$ 
it suffices to retain relatively small number of terms, as the 
stabilization of the approximants around exact frequency $\omega$ 
is quite fast. However, for the low-lying fundamental modes and 
their overtones the stabilization of the results may be slow. 
Indeed, the approximants may be scattered on the complex plane without 
any visible pattern or trend, even for a great number of terms
retained in the series. However, in many cases, starting with 
some $N',$ the Pad\'e transforms start to stabilize and taking 
into account additional terms leads to the improvement of the result. 
Unfortunately, there are cases in which $N'$ is either too big to 
calculate the quasinormal modes in a reasonable time or it does 
not exist at all. Here we shall adopt a pragmatic point of view: 
If $N' >4k,$ where $k$ is the maximal order of the Pad\'e approximant 
(in this paper $k = 300$), then we classify the mode as the mode 
that cannot be calculated within the WKB-Pad\'e framework. It does 
not mean that taking larger $k$ would not improve the results. 
However, it should be emphasized that increasing $k$ may be impractical 
or even (due to limited computational resources or the 
very nature of the problem) impossible.

The frequencies of the fundamental quasinormal mode $\ell=2$ and its 
few lowest overtones are tabulated in Tab.~\ref{tab4a}. Both the 
continued fractions method and the Hill determinant method agree 
to (at least) the quoted accuracy. Unfortunately, the slow convergence of the
WKB-Pad\'e approximants for $n =0,$  and their  apparent lack of stabilization 
for higher overtones suggest that the WKB-Pad\'e method does not work in this case. 
Indeed, for $\ell =2,\, n =0$ mode it yields $\tilde{\omega} = 7.15761193 - 2.11830698 i$ 
and  $\Delta_{R}\omega \approx 0.5\%$ and  $\Delta_{I}\omega \approx 2.7  \%,$ 
which is a poor result. 

To examine this discrepancy we introduce the Borel-Le Roy summation~\cite{le}. 
Let $S = \sum_{k} f_{k} \varepsilon^{k} $ be the asymptotic
series expansion of some function $f(\varepsilon).$ Now, define 
a new series by dividing each term by a factor $\Gamma(k+b+ 1)$
\begin{equation} 
 \mathcal{L}^{b}(\varepsilon) = \sum_{k=0}^{\infty}\frac{f_{k}}{\Gamma(k+b+1)} \varepsilon^{k},
 \label{bl}
\end{equation}
where $b$ $(\Re (b) > -1)$ is some adjustable parameter.
Since
\begin{equation}
 \Gamma(k+b+1) = \int_{0}^{\infty} dt \,e^{-t} t^{k+b}
\end{equation}
one can use this equation to `reinsert' the $\Gamma$ factor as follows
\begin{equation}
 f(\varepsilon) = \int_{0}^{\infty} dt \, e^{-t} t^{b} \mathcal{L}^{b} (\varepsilon t).
\end{equation}
Unfortunately, we do not know how to sum the series (\ref{bl}) and if only a finite number of terms
is known (and this is the case here) then the integral gives the original series $S$.
The essence of the method is to apply the Pad\'e summation to the truncated series (\ref{bl})
\begin{equation}
  S_{BLR}(\varepsilon) = \int_{0}^{\infty} dt \, e^{-t} t^{b} \mathcal{P}^{M}_{N} (\varepsilon t),
\end{equation}
where $\mathcal{P}^{M}_{N}$ is the Pad\'e transform, and to
calculate the integral numerically. Putting $b=0$ leads to the Borel summation~\cite{CarlB}. 
The Borel summation
of the expansion (\ref{omm}) has been introduced by Hatsuda in Ref.~\cite{hatsuda} (see also~\cite{jago,Eni}).
It should be noted that the integration may be time consuming and numerically unstable.

Now, to minimize the danger of numerical instabilities,
we calculate the quasinormal frequency of the $(j=2, l=2,n=0)$ mode
using the Borel-Le Roy summation for $b = 0, 1$ and $2$.
Quite interestingly, $\omega$ calculated using this technique
agrees with the frequency calculated using the WKB-Pad\'e method. 
Although both WKB-based methods 
give the same frequency, we believe that its correct value is given 
by the Hill determinant method and the continued fraction method.
On the other hand however, inspection of Tables~\ref{tab4} and~\ref{tab5}  
shows that for $\ell \geq 3$ we have a very good agreement of the results 
obtained within the frameworks of the all three methods. All this suggests 
that the WKB-Pad\'e summation is responsible for this discrepancy.
It is unclear if retaining more terms in Eq.~(\ref{omm}) would improve
the WKB result. May be some radical modifications of the method are needed.

Once again, each result presented in the second column of Tables~\ref{tab4a}-\ref{tab5}
has been calculated using four methods: two
independent calculations making use of the determinant of the four-diagonal and the tridiagonal
matrices and two independent calculations of the continued fractions making 
use of Eqs. (\ref{cont2}) and (\ref{cont3}),
respectively. We believe that this minimizes the danger of some accidental errors.

For the mode $\ell = 3$ and its 
four first overtones the accuracy of the WKB-Pad\'e based calculations is better than $9.8 \times 10^{-6}\%$ 
and $3.7 \times 10^{-5} \%$ for the real and the imaginary part of the frequencies, respectively.
For $\ell = 4$ we have a perfect agreement of all three methods.

\begin{center}
\begin{table}
 \caption{\label{tab4a} 
The frequencies of the quasinormal modes of the
gravitational vector perturbations $(\ell = 2)$
of the five-dimensional Schwarzschild-Tangherlini black hole calculated for 
$n = 0,1,2,3,4.$ The Hill determinant method (HD), the continued-fractions
method (CF), both with the convergence acceleration yield identical results. 
The frequencies are defined as $\tilde{\omega} = \omega/T_{H},$
where the Hawking temperature $T_{H} = 1/ 2 \pi.$ }
\begin{tabular}{|c | c| }
\hline\hline
$ n$ & $\tilde{\omega}_{CF}/\tilde{\omega}_{HD}$  \\ \hline
 0 & $   7.12515163375-2.05788530886 i $\\
 1 & $   5.95278904708-6.42166605748 i $\\
 2 & $  3.41133340871-12.09301936730 i $\\
 3 & $  2.73756226817-19.60939205419 i $\\
 4 & $  2.51060685222-26.26273928972 i $\\
\hline\hline
\end{tabular}
\end{table}
\end{center}

\begin{center}
\begin{table}
 \caption{\label{tab4} 
 The frequencies of the quasinormal modes of the 
 gravitational vector perturbations $(\ell = 3)$
of the five-dimensional Schwarzschild-Tangherlini black hole calculated for 
$n = 0,1,2,3,4.$ The Hill determinant method (HD) and the continued-fractions
method (CF), both with the convergence acceleration,  yield identical results. 
The WKB-Pad\'e results are slightly less accurate. 
The frequencies are defined as $\tilde{\omega} = \omega/T_{H},$
where the Hawking temperature $T_{H} =1/ 2 \pi,$ and the last column gives 
the maximal order of the (diagonal) Pad\'e approximants.
 }
\begin{tabular}{|c| c | c| c| }
\hline\hline
$n$ & $ \tilde{\omega}_{CF}/\tilde{\omega}_{HD}$  & $\tilde{\omega}_{WKB}$ & Pad\'e  \\ \hline
     0 & $     \underline{10.84077672458}  - \underline{2.09758650296} i $ & $  \underline{10.84077672458}   - \underline{ 2.09758650296} i $ &    (200,200) \\
     1 & $     \underline{10.16678579574}  - \underline{6.40845617817} i $ & $     \underline{ 10.16678579574}   - \underline{ 6.40845617817} i  $ &   (200,200)  \\
     2 & $      \underline{8.88194273}632  - \underline{11.09290342}86 i $ & $      \underline{8.88194273}749    - \underline{ 11.09290342}781 i  $ &   (200,200)  \\
     3 & $      \underline{7.216020}42842  - \underline{16.397974}9156 i $ & $      \underline{7.216020}52137    - \underline{ 16.397974}79356 i  $ &   (250,250)  \\
     4 & $      \underline{5.527809}70687  - \underline{22.3321}576615 i $ & $      \underline{5.527809}16599    -  \underline{22.3321}4931620 i  $ &   (300,300)  \\
\hline\hline
\end{tabular}
\end{table}
\end{center}

\begin{center}
\begin{table}
 \caption{\label{tab5} 
 The frequencies of the quasinormal modes of the  
gravitational vector perturbations $(\ell = 4)$
of the five-dimensional Schwarzschild-Tangherlini black hole calculated for 
$n = 0,1,2,3,4.$ The Hill determinant method (HD), the continued-fractions
method (CF), both with the convergence acceleration,  
and the WKB-Pad\'e method yield identical results. 
The frequencies are defined as $\tilde{\omega} = \omega/T_{H},$
where the Hawking temperature $T_{H} =1/ 2 \pi,$ and the last column 
gives the maximal order of the (diagonal) Pad\'e approximants.
 }
\begin{tabular}{|c| c | c| c| }
\hline\hline
$n$ & $\tilde{\omega}_{CF}/\tilde{\omega}_{HD}$   & $\omega_{WKB}$ & Pad\'e  \\ \hline
 0    & $     \underline{ 14.328729987193} - \underline{2.136391879254} i      $ & $        \underline{14.328729987193} -  \underline{2.136391879254} i     $ &     (150,150) \\
 1    & $       \underline{13.825163344184} -  \underline{6.482259700362} i      $ & $        \underline{13.825163344184} -  \underline{6.482259700362} i     $ &     (200,200) \\
 2    & $      \ \underline{12.850614230172} -  \underline{11.057416573784} i     $ & $       \underline{12.850614230172} -  \underline{11.057416573784} i    $ &     (200,200) \\
 3    & $       \underline{11.506972273038} -  \underline{16.027390914015} i     $ & $       \underline{11.506972273038} -  \underline{16.027390914015} i    $ &     (200,200) \\
 4    & $        \underline{9.998966688866} -  \underline{21.500865800593} i     $ & $        \underline{ 9.998966688866} -  \underline{21.500865800593} i    $ &     (200,200) \\
\hline\hline
\end{tabular}
\end{table}
\end{center}

\begin{figure}
\centering
\includegraphics[width=11cm]{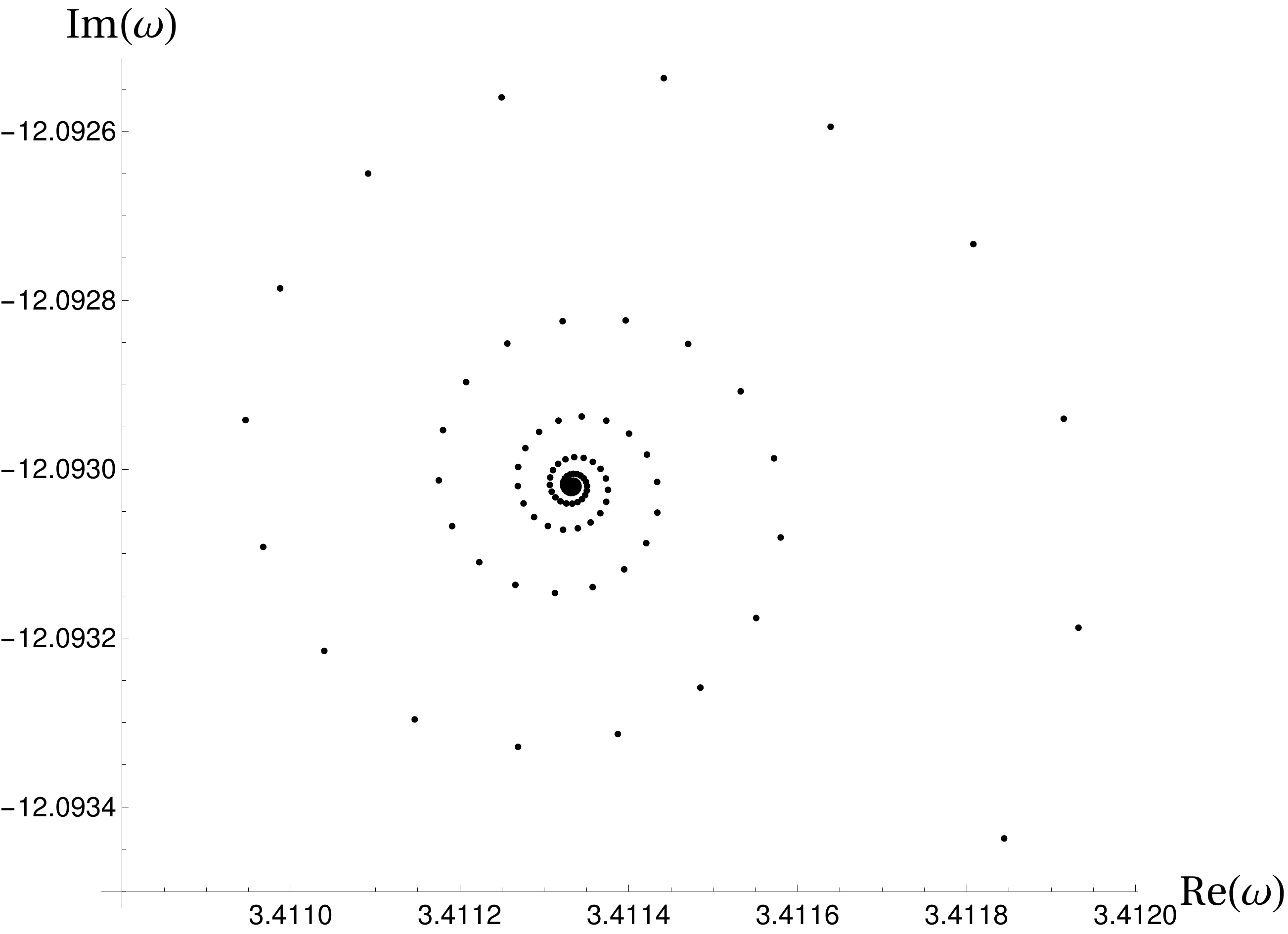}
\caption{The gravitational vector perturbations $(j = 2)$. Migration of the approximants 
of the complex frequency of the $\ell =2, n =2$  mode on the complex plane.
The red dot represents the limiting value calculated using the Wynn acceleration. 
}
\label{fig1}
\end{figure}

\begin{figure}
\centering
\includegraphics[width=11cm]{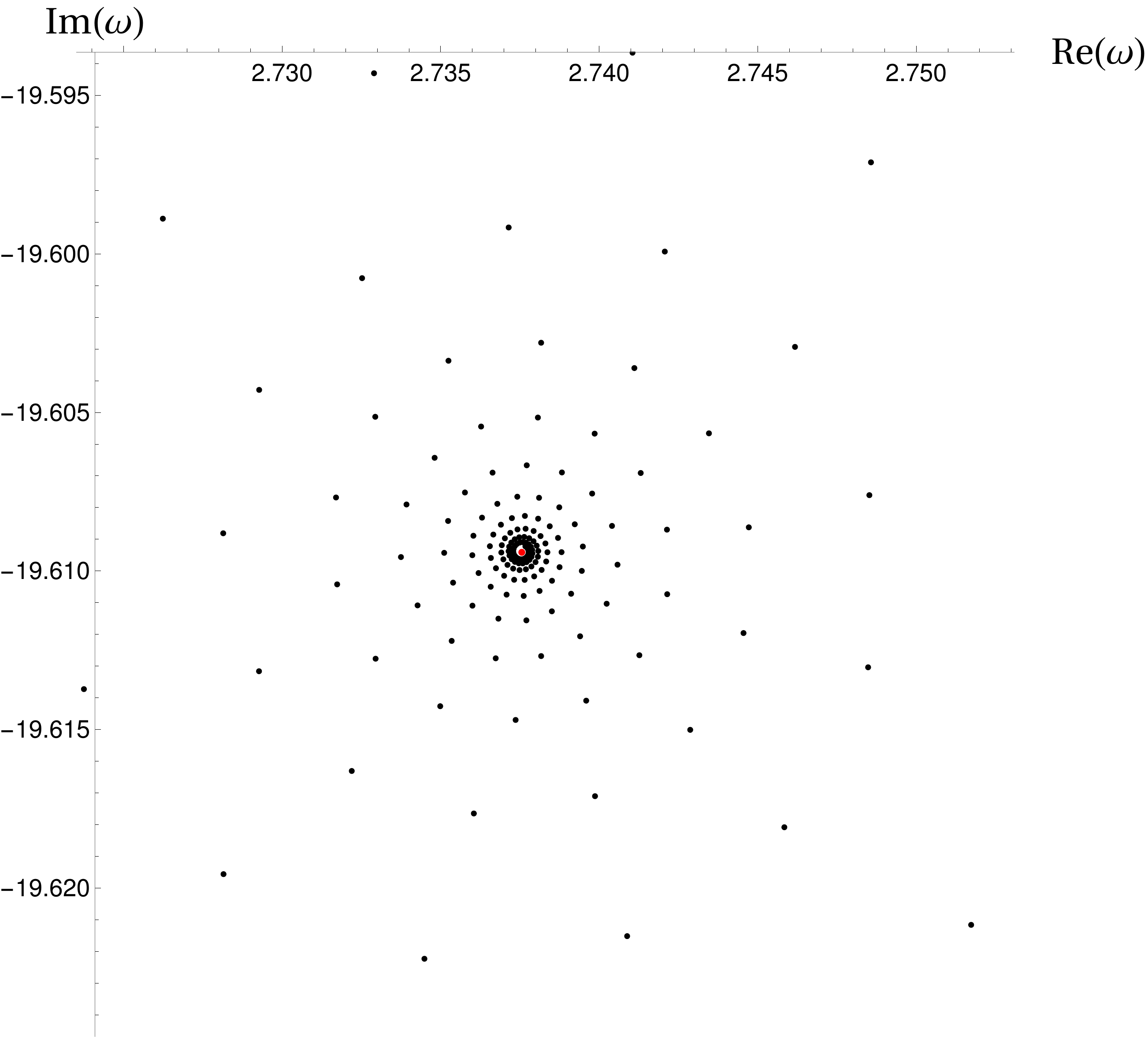}
\caption{The gravitational vector perturbations $(j = 2)$. Migration of the approximants 
of the complex frequency of the $\ell =2, n =3$  mode on the complex plane.
The red dot represents the limiting value  calculated using the Wynn acceleration. 
}
\label{fig2}
\end{figure}

\begin{figure}
\centering
\includegraphics[width=11cm]{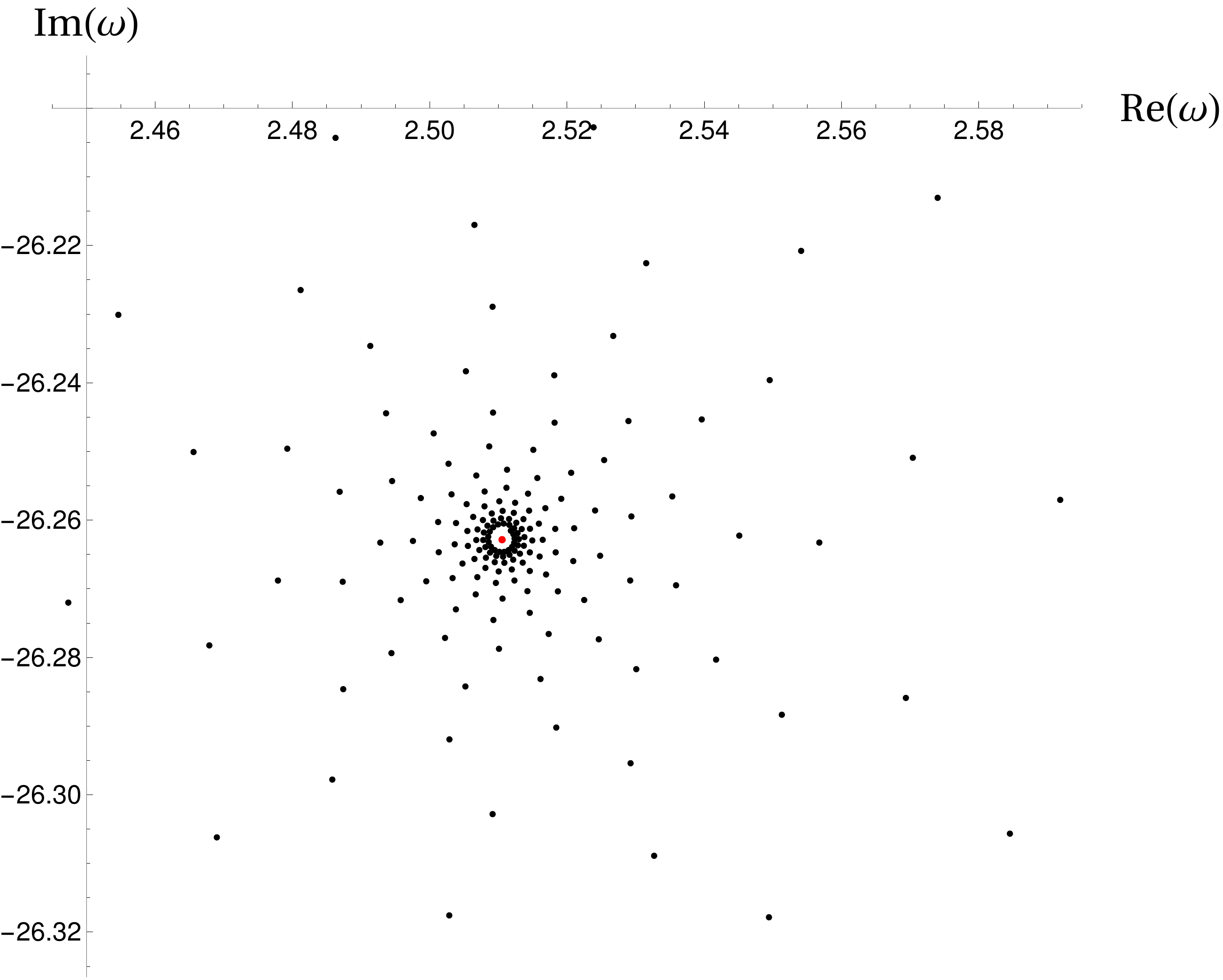}
\caption{The gravitational vector perturbations $(j = 2)$. Migration of the approximants 
of the complex frequency of the $\ell =2, n =4$  mode on the complex plane.
The red dot represents the limiting value calculated using the Wynn acceleration.
}
\label{fig3}
\end{figure}

It is of some interest to analyze how the roots of the polynomials 
$p_{n}$ (see Sec.~\ref{HDm} and~\ref{CFm} ) identified as the consecutive 
approximations of the quasinormal frequencies migrate on the complex 
plane. For the perturbations considered in this paper it depends 
on the number of the overtone as follows. Frequency of the fundamental 
modes and its  first overtone $(n =1)$ rapidly approach their limiting 
values. On the other hand, starting with $n =2$ the approximate frequencies 
lie on the spiral curve, with the exact value at its center. This behavior 
is clearly visible in Figs.~\ref{fig1}-\ref{fig3} and the red dot at the center 
represents the result of the application of the $\epsilon$-acceleration 
to the series of approximate roots. This suggests that the final result 
is encoded in relatively small number of approximants. Moreover, increasing 
$n$ while keeping $N$ fixed  places the approximants on the complex plane  
farther from the exact value and to sustain accuracy one has to increase $N.$
This remark concerns more the Hill determinant method than the method of continued fraction 
as in the latter case there is a simple way to estimate the contribution of the remaining part
of the continued fractions.

Putting $j =2/3$ in Eq.~(\ref{ma_eq})  results in the equation considered previously by 
Crispino, Higuchi, and Matsas (see Eqs.~2.42 and 2.48 in Ref.~\cite{Crispino}).
Inspection of the Tables~\ref{tab6}-\ref{tab8} shows  that the results constructed 
using all three methods are in perfect agreement.   
The frequencies given by the WKB-Pad\'e method are either exactly the same as the ones calculated 
using  the continued fraction method and the Hill determinant method
or slightly less accurate. For example, for $\ell = 1$ and $n =4$ (which is the  
hardest case considered in this paper for the electromagnetic vector perturbations) one has 
$\Delta_{R}\omega \approx 0.57\times 10^{-5}\%$ and  $\Delta_{I}\omega \approx 0,11\times 10^{-5}\%.$ 
And this is still amazingly accurate result. The WKB-Pad\'e results can be made even more accurate
simply by retaining additional terms in the expansion (\ref{omm}).

\begin{center}
\begin{table}
 \caption{\label{tab6}  
  The frequencies of the quasinormal modes of the electromagnetic vector perturbations 
  $(\ell = 1)$
of the five-dimensional Schwarzschild-Tangherlini black hole calculated for 
$n = 0,1,2,3,4.$ The Hill determinant method (HD) and the continued-fractions
method (CF), both with the convergence acceleration,  yield identical results. 
The WKB-Pad\'e results are slightly less accurate. The frequencies are defined as 
$\tilde{\omega} = \omega/T_{H},$
where the Hawking temperature $T_{H} = 1/2 \pi,$ and the last column gives 
the maximal order of the (diagonal) Pad\'e approximants.
 }
\begin{tabular}{|c| c | c| c| }
\hline\hline
$n$ &  $ \tilde{\omega}_{CF}/\tilde{\omega}_{HD}$  & $\tilde{\omega}_{WKB}$ & Pad\'e  \\ \hline
 0  &$   \underline{ 5.98616712253} -  \ \underline{2.20376124623} i  $&$        \underline{5.98616712253}  -  \ \underline{2.20376124623} i  $&  (100,100)  \\             
 1  &$    \underline{4.93596110359} -   \underline{7.06760374979} i  $&$       \ \underline{4.93596110359}  -  \ \underline{7.06760374979} i  $&  (100,100)  \\
 2  &$   \underline{ 3.65875874}531 -  \underline{12.958071585}91 i  $&$        \underline{3.65875874}456  -  \underline{12.958071585}56 i  $&  (150,150)  \\
 3  &$    \underline{2.836146}54139 -  \underline{19.342239}30964 i  $&$        \underline{2.836146}40894  -  \underline{19.342239}20621 i  $&  (200,200)  \\
 4  &$    \underline{2.33220}463436 -  \underline{25.78606}028998 i  $&$        \underline{2.33220}330869  -  \underline{25.78606}310021 i  $&  (250,250)  \\
\hline\hline
\end{tabular}
\end{table}
\end{center}

\begin{center}
\begin{table}
 \caption{\label{tab7}  
  The frequencies of the quasinormal modes of the gravitational vector perturbations 
  $(\ell = 2)$
of the five-dimensional Schwarzschild-Tangherlini black hole calculated for 
$n = 0,1,2,3,4.$ The Hill determinant method (HD) and the continued-fractions
method (CF), both with the convergence acceleration,  yield identical results. 
The WKB-Pad\'e results are almost as accurate as the previous two. The frequencies 
are defined as $\tilde{\omega} = \omega/T_{H},$
where the Hawking temperature $T_{H} = 1/2 \pi,$ and the last column gives the maximal 
order of the (diagonal) Pad\'e approximants.
 }
\begin{tabular}{|c| c | c| c| }
\hline\hline
$n$ & $ \tilde{\omega}_{CF}/\tilde{\omega}_{HD}$  &  $\tilde{\omega}_{WKB}$ & Pad\'e  \\ \hline 
    0     &$    \underline{9.2271133280308}  -    \underline{2.2143549304187} i  $&$   \underline{9.2271133280308} -    \underline{2.2143549304187} i  $& (100,100) \\
    1     &$    \underline{8.4728284656951}  -   \underline{ 6.8458616593904} i   $&$   \underline{8.4728284656951} -    \underline{6.8458616593904} i  $& (100,100) \\
    2     &$    \underline{7.1966403739904}  -   \underline{12.0804073774333} i   $&$   \underline{7.1966403739904} -   \underline{12.0804073774333} i  $& (150,150) \\
    3     &$    \underline{5.9189890960475}  -  \underline{1 8.034284652101}4 i   $&$   \underline{5.9189890960475} -   \underline{18.034284652101}5 i  $& (150,150) \\
    4     &$   \underline{ 4.968250934553}7  -   \underline{24.383217919758}9 i   $&$   \underline{4.968250934553}5 -  \underline{ 24.383217919758}5 i  $& (250,250) \\
\hline\hline
\end{tabular}
\end{table}
\end{center}

\begin{center}
\begin{table}
 \caption{\label{tab8} 
The frequencies of the quasinormal modes of the  
electromagnetic vector perturbations $(\ell = 3)$
of the five-dimensional Schwarzschild-Tangherlini black hole calculated for 
$n = 0,1,2,3,4.$ The Hill determinant method (HD), the continued-fractions
method (CF), both with the convergence acceleration,  and the WKB-Pad\'e 
method yield identical results. 
The frequencies are defined as $\tilde{\omega} = \omega/T_{H},$
where the Hawking temperature $T_{H} = 1/ 2 \pi,$ and the last column gives 
the maximal order of the (diagonal) Pad\'e approximants.
}
\begin{tabular}{|c| c | c| c| }
\hline\hline
$n$ &  $ \tilde{\omega}_{CF}/\tilde{\omega}_{HD}$  & $\tilde{\omega}_{WKB}$ & Pad\'e  \\ \hline
   0   &$    \underline{12.4184186551775}  -  \underline{ 2.2177094497067} i     $&$     \underline{12.4184186551775}  -   \underline{2.2177094497067} i     $&      (100,100) \\
   1   &$    \underline{11.8412014398195}  -   \underline{6.7660501645214} i     $&$     \underline{11.8412014398195}  -   \underline{6.7660501645214} i     $&      (150,150) \\
   2   &$    \underline{10.7694063181563}  -  \underline{11.6625811156602} i     $&$     \underline{10.7694063181563}  -  \underline{11.6625811156602} i     $&      (150,150) \\
   3   &$     \underline{9.4315982617932}  -  \underline{17.1019392490840} i     $&$      \underline{9.4315982617932}  -  \underline{17.1019392490840} i     $&      (200,200) \\
   4   &$     \underline{8.1520905535350}  -  \underline{23.0769667314609} i     $&$      \underline{8.1520905535350}  -  \underline{23.0769667314609} i     $&      (200,200) \\
\hline\hline
\end{tabular}
\end{table}
\end{center}

\section{Discussion and Final Remarks}
\label{final}

All perturbations considered so far are described (in a linear regime) 
by the equation (\ref{master}), and, consequently, the coefficients
$a_{k}$ of the series expansion (\ref{psi_odd}) satisfy the four-term 
recurrence relation (\ref{rec_1}). As the continued fraction method
requires a three-term recurrence, the relations (\ref{rec_1}) should 
be transformed to the required form by the Gauss elimination. 
Since we are interested in the low-lying fundamental modes and 
their overtones satisfying $(n \leq 4)$ we employed the series 
acceleration technique. On the other hand, 
in the Hill-determinant method one can use both the four-term or 
the three-term recurrence.  The determinant of  $(n+1\times n+1)$ matrix 
constructed form the infinite sparse band matrix of width 4 can 
be calculated efficiently with the aid our formula (\ref{rec_hill_1}). 
Similarly, one can calculate the determinant of the tridiagonal matrix. 
A typical migration of the roots that approximate quasinormal frequency 
on the complex plane is shown in Figs.~\ref{fig1}-\ref{fig3}. For 
the low-lying overtones, the roots approach their limiting value quite fast,
whereas the general trend for higher overtones suggests that     
the convergence becomes slower and slower.
To secure great accuracy of the quasinormal frequencies for the higher 
overtone numbers, the dimension of the matrices should be unreasonably 
big and to accelerate convergence we have used, once again,  the Wynn's $\epsilon$-method.
On the other hand, the gravitational scalar perturbations described 
by the potential (\ref{grav_scalar}) lead to the eight-term recurrence. 
Since (\ref{grav_scalar}) does not belong to the class
of potentials considered in this paper, here we report only on our preliminary
results obtained within the framework of the WKB-Pad\'e approximation.
The tables of the quasinormal frequencies (to 4 decimal places) calculated
using the method of continued fractions are given in Ref.~\cite{jose_5_dim2}.
Our results for $\ell = 2$ (when transformed to the normalization adopted in~\cite{jose_5_dim2})  
are  $ 0.94774 - 0.25609 i, 0.85123 - 0.82116 i, 0.67274 - 1.54307i$ and  
$0.50889 - 2.43311 i$ for $ n =0,1,2,$ and $3$, respectively, and when 
rounded to four decimal places they are exactly the same as those 
presented in Ref.~\cite{jose_5_dim2}. 
This shows the power of the WKB-Pad\'e method and simultaneously confirms 
correctness of the results obtained by Cardoso, Lemos and Yoshida.
It should be noted, that because of a complicated form 
of the gravitational scalar potential, the calculations are much more 
involved. 

Let us return to the Hill determinant method applied to perturbations leading 
to $k$-term recurrences $(k > 3).$ As has been explained earlier, one can use 
either the Gauss elimination to obtain the tridiagonal matrix or work with 
the original recurrence. To calculate determinants in the latter case efficiently,  
one should generalize Eq.~(\ref{rec_hill_1}). In real calculations the computational 
complexity of each method should be estimated, i.e, the following question  
should be asked. What is more time-consuming: Calculations of determinants 
of banded matrices of width $k$ with relatively simple matrix elements, 
or performing $(k-3)$ consecutive Gauss eliminations and  calculating 
determinants of tridiagonal matrices with quite complicated elements?

It should be emphasized that the perfect agreement between the results 
obtained using the Hill determinant method and the method of continued 
fractions, although expected, is quite impressive and the role of the 
convergence acceleration should not be underestimated in this regard.
On the other hand, the performance of the WKB-Pad\'e method (within domain 
of its applicability) is really amazing. In most cases it is better 
than any competing WKB-based technique. Of course,  such a comparison 
 is somewhat unfair given the much greater complexity of our method.

The techniques presented in this paper can easily be adapted to other 
dimensions. However, there are a few bottlenecks that may pose a challenge 
to the calculations of the quasinormal modes. For the method of continued 
fractions and the Hill determinant method it would be a complexity 
of the equations which have to be solved. At any order one has to identify 
all solutions and this may be a time-consuming process. On the other hand, 
the real problem of the WKB-Pad\'e method is the necessity to calculate 
the consecutive derivatives of the potential at its maximum. Since 
the calculations are carried out analytically it places severe demand 
on the computer resources, especially for more complex potentials.
Moreover, it turns out that the time spent on construction of the Pad\'e transforms 
is only a small fraction of the total time of computations and for a given $N,$ the 
calculation time of the WKB series is practically insensitive to the type 
of the black hole perturbation.

Finally, let us briefly discuss the strength and limitations of each 
method employed in this paper. Of course, their great positive aspect 
is the ability to construct the highly accurate values of $\omega.$
However, they lose accuracy with increasing the overtone number, 
$n,$ even though we use the convergence acceleration algorithms. 
For the continued fraction method and the method of the Hill determinant 
the deterioration of the results is slow, and to secure assumed
accuracy it is sufficient to increase the number of terms retained 
in the expansion~(\ref{psi_odd}). For highly damped modes this may 
be impractical or insufficient. In such a case it would be reasonable
to construct the asymptotic approximation of the remainder of the continued 
fraction and to modify the root searching algorithm. On the other hand, 
the WKB-Pad\'e method has limited applicability for overtones 
of the low lying modes, which is reflected in slow stabilization (if any)
of the series of approximants. This can be deduced form the trend
clearly visible in the tables. It is not clear if increasing the number 
of terms in (\ref{omm}) would cure the problem. Such calculations 
are both demanding and time consuming and we intend to return to 
this group of problems elsewhere.


\end{document}